\documentstyle[preprint,aps]{revtex}
\begin{document}
\preprint{WIS-94/48 Nov-PPH}
\newcommand{\bms}{{\mbox{\boldmath $s$}}}
\newcommand{\bmP}{{\mbox{\boldmath $P$}}}
\newcommand{\bmp}{{\mbox{\boldmath $p$}}}
\newcommand{\bmR}{{\mbox{\boldmath $R$}}}
\newcommand{\bmr}{{\mbox{\boldmath $r$}}}
\newcommand{\bmz}{{\mbox{\boldmath $r_1$}}}
\newcommand{\bmzp}{{\mbox{\boldmath $r_1'$}}}
\newcommand{\bmrp}{{\mbox{\boldmath $r'$}}}
\newcommand{\bmq}{{\mbox{\boldmath $q$}}}
\newcommand{\bmb}{{\mbox{\boldmath $b$}}}
\newcommand{\NA}{{\rm N/A}}
\newcommand{\alN}{{\alpha/{\rm N}}}
\newcommand{\alA}{{\alpha/{\rm A}}}

\draft
\date{\today}
\title{A quark-cluster description of the EMC effect\footnote
{Seminar given in the 18th Course of the "International
School on Nuclear Physics", 16-23/9/'94, Erice, IT. To
be published in
Progress in Particle and Nuclear Physics, Vol. 34, Pergamon Press.}}
\author{S. A. Gurvitz and A.S. Rinat}
\address{Department of Particle Physics, Weizmann Institute of
         Science, Rehovot 76100, Israel}
\maketitle
\begin{abstract}
The response  of a nucleus  composed of  nucleons and confined  quarks is
calculated  in a  non-relativistic cluster  model and  is shown  to be  a
convolution  of responses  of, respectively,  a nucleus  with interacting
point-nucleons and of a nucleon composed of quarks.  We further show that
the apparent off-shell energy in the expression for the nucleon structure
function in Impulse Approximation does not constitue a real problem.
We conjecture a generalization to the  relativistic regime.
\end{abstract}
\section{Introduction}
Consider inclusive  scattering of an  unpolarized lepton from  a nucleus
$'A'$ under the assumption
that  the  cross  section  can  be factored  into  an
$eN$ Mott cross section and a structure function $F_{\alA}(x,Q^2)$.
The kinematic variables are $Q^2=-q^2$ and
$x=Q^2/2Pq$, the Bjorken variable for target 4-momentum $P$.
The EMC  effect is then the  observation that $F_{\alA}\ne AF_{\alN}$,
i.e. the structure function per nucleon of a nucleus made up of quarks
($'\alpha'$) is not the averaged nucleon structure function \cite{aub}

Calculations  of  $F_{\alA}$  are usually based on the Plane Wave Impulse
Approximation  (PWIA) in which  the  nucleus  provides a  nucleon,  the
constituents of  which subsequently  absorb $q$ (Fig. 1).  The expression
in the $Q^2\to\infty$  limit, usually cited in the literature
\cite{btom}.
\begin{eqnarray}
F_{\alA}^{PWIA}(x)=A^{-1}\int dz h_{\NA}(z)F_{\alN}(x/z),
\label{a1}
\end{eqnarray}
where $z$ is the light-cone momentum fraction of a nucleon in
the nucleus.
The  above  expression  appears  as  a  generalized  convolution  of  two
dynamic quantities.  One is some distribution function $h$ of a nucleon
in a  target while
the  second  element is  the  structure  function  of a  nucleon.   Those
components  refer,  respectively, to  a  system  of point-nucleons  in  a
nucleus without reference  to quarks, and to the structure  function of a
nucleon composed of quarks, without a trace  that the former is part of a
nucleus.  Differently stated, the expression   (1) seems to be compatible
with the dynamics of a cluster  model with independent degrees of freedom
for quarks in a nucleon and nucleons in a nucleus.  It is then natural to
exploit a cluster  model for a study  of the EMC effect, not  only in the
PWIA (essentially the $Q^2\to\infty$ limit), but also for general, finite
$Q^2$. This implies retention of  Final State Interactions (FSI) which in
principle involve (hadronized) quarks,  gluons, etc.  Of course,
without introducing  drastic simplifications, there is little hope for a
realistic treatment.

In the  following we shall formulate the simplest conceivable model,
namely a non-relativistic (NR) quark cluster model, which shall
provide a formally exact expression for $F_{\alA}$.
It also permits proper
handling of a recurring off-shell problem, encountered in realistic
approaches. In the end we suggest an extension  of the model to
the relativistic regime.

\section{The total nuclear response in a quark cluster model.}

Consider 3$A$ spin-0 quarks of equal charge, with triplets clustering
in bags. No details will be required regarding the
dynamics  of quarks inside one  bag.
In  contrast one has to be specific about the interaction between quarks
belonging to different bags. Instead of those, we consider the total
interactions between two triplets of quarks. Those we replace
by  an effective,  $'$optical$'\ NN$  potential, acting between the
centers of the bags, thereby violating in part the Pauli principle.
The above potential ought  to be complex and energy-dependent
but we simply take  it to  be real and  energy-independent.
Rather than seeking justification for
a cluster model (bag excitation energies  which  much exceed
those of a nucleus, etc.) we explore its predictions.

We thus assume  quarks to move in  a confining potential
$V(r_{\ell,i})$, where ${\bmr}_{\ell,i}$  are quark coordinates
relative to ${\bmR}_i$. The latter is the center-of-mass coordinate of
nucleon $i$  with respect to the  center-of-mass of the nucleus.
The   total  nuclear
potential is  assumed to  be $U=\sum_{j<i}U_{ij}(R_{ij})$,  where
$R_{ij}=|{\bmR}_i -{\bmR}_j|$. The cluster Hamiltonian is then
\begin{eqnarray}
H=\sum_{i=1}^A \left(\sum_{\ell =1}^3
\left(-\frac{1}{2m}\nabla^2_{\ell ,i}+ V(r_{\ell ,i})\right)
-\frac{1}{2M} \nabla^2_{i}+\sum_{j<i}U(R_{ij})\right),
\label{a2}
\end{eqnarray}
with $m$  and $M$ the  quark and nucleon mass. By choice
the total nuclear potential is independent of the
relative quark coordinates $\bmr$, and so is the quark
part  of  nuclear  coordinates.
States $|\Psi\rangle$  consequently factor into internal nucleon
and nuclear states $|n\rangle,|N\rangle$ with respective
energies $e_n,E_N$.

We now consider the NR structure function or response $W_{\alA}$ of
the composed nucleus ($'$the total nuclear response$'$) as function of
$q=(\nu,\bmq)$ and concentrate on the incoherent part which dominates
for large $q$. Its contribution per nucleon is
\begin{eqnarray}
W_{\alpha/{\rm A}}(q,\nu )
=3\sum_n\sum_N {\hspace {-0.5cm}{\int }}\,\,|\langle 0,0|
\exp \left[ i\bmq (\bmr_{11} +
\bmR_1)\right] |n,N \rangle|^2\delta
(\nu -E_{N0}-e_{n0}-q^2/2M_{\rm A}),
\label{a3}
\end{eqnarray}
where $E_{N0}=E_N-E_0;\, e_{n0}=e_n-e_0$ are internal excitation
energies. The spectrum  of confined
quarks is discrete, while nuclear states $|N\rangle$ may
belong to either discrete or to continuum parts of a spectrum.

The form of the charge-density operator in (\ref{a3}) permits
a separate treatment of parts related to, respectively nucleons in a
nucleus  and quarks in a nucleon. We start with a remark on the latter.
Although a  free nucleon, at rest or moving, is
always on its energy shell, a constituent one
in a nuclear medium has a non-vanishing momentum  $\bmP$ and is
generally off-shell with
${\cal E}\ne {\cal E}_{\bmP}={\bmP}^2/2M$. Limiting ourselves to
the case of an on-shell  recoiling nucleon with momentum $\bmp+\bmq$,
we have the general expression
\begin{mathletters}
\label{a4}
\begin{eqnarray}
W_{\alN}(q,\nu;\bmP,{\cal E})&=&3\sum_{n}\left |\langle 0|\exp\left
(i\bmq \bmr_{11}\right )|n\rangle\right |^2
\delta \bigg (\nu+{\cal E} -e_{n0}-{\cal E}_{\bmP+\bmq}\bigg )
\label{a4a}\\
&=&W_{\alN}(q,\nu+{\cal E}-{\cal E}_{\bmP+\bmq}+{\cal E}_{\bmq})
\label{a4b}
\end{eqnarray}
\end{mathletters}
Eq. (\ref{a4b}) shows that  one can express a NR structure function for
a moving target with off-shell kinematics  in terms of the same for
an on-shell target at rest $W(q,\nu)=W(q,\nu;0,0)$ by a mere shift
in the energy argument.

Next we consider the response per nucleon of a nucleus of $A$ point-like
nucleons (in short $'$the nuclear response$'$)
\begin{eqnarray}
W_{{\rm N/A}}(q,\nu )=\sum_{N} {\hspace {-0.5cm}{\int }}\,\,
\left |\langle 0|\exp\left
(i\bmq\bmR_1 \right )|N\rangle
\right |^2\delta (\nu-E_{N0}-q^2/2M_{\rm A})
\label{a5}
\end{eqnarray}
{}From Eqs. (\ref{a3}), (\ref{a4}) and (\ref{a5}) one then easily finds
the desired response of the
system under consideration in terms of the  same for a free nucleon and a
nucleus composed of point  nucleons
\begin{eqnarray}
W_{\alA}(q,\nu)
=\int^{\nu}_{q^2/2M_A} d\nu'\,W_{\NA}(q,\nu')
W_{\alN}(q,\nu-\nu'-{\bmq}^2/2M)
\label{a6}
\end{eqnarray}
The  above exact,  non-perturbative expression  refers only  to component
structure  functions for  on-shell  kinematics.  Note  that  in spite  of
expectations,  there  is  no  implicit  evidence  for  the  motion  of  a
constituent nucleon in a nucleus. We shell return to this point in
Section  III.

In the following it will be useful to replace the energy transfer $\nu$
by the West variable $y$
\cite{grs,wes}
\begin{eqnarray}
y=-\frac{q}{2}+\frac{m\nu}{q},
\label{a7}
\end{eqnarray}
which is  the minimal momentum of the struck particle in the direction of
$\hat \bmq$. Parallel to $y$ above for the nucleon response, one
defines a scaling variable appropriate to $W_{\NA}$
\begin{eqnarray}
Y= -\frac{q}{2}+\frac{M\nu}{q}=\frac{M}{m}\left (y+\frac{M-m}{2M}q\right)
\label{a8}
\end{eqnarray}
In terms of those one defines  reduced response, for instance for a
nucleon
\begin{eqnarray}
\phi_{\alN}(q,y)=(q/m)W_{\alN}(q,\nu)
\label{a9}
\end{eqnarray}
One then readily  checks that (\ref{a6}) becomes
\begin{eqnarray}
\phi_{\alpha/{\rm A}} (q,y)=\int_{Y_{min}}^{Y_{max}}\, dY
\phi_{{\rm N/A}}(q,Y)\phi_{\alpha/{\rm N}}
\left(q,y-\frac{m}{M}Y\right),
\label{a10}
\end{eqnarray}
where $Y_{min}=-(q/2)(1-M/M_A)$ and
$Y_{max}=(M/m)y+(q/2)(M/m-1)$.

At  this point  we emphasize  that  Eqs. (\ref{a6})  and (\ref{a10})  are
formally  exact expressions for the  desired response in
the cluster  model and  are valid  not only  asymptotically, but  for any
finite $q$.  They have the form of a convolution, involving on equal
footing nucleon  and nuclear responses $W_{\alpha/{\rm  N}}$ and $W_{{\rm
N/A}}$.  Separability  of degrees of freedom  in a NR model  is clearly a
sufficient, and quite likely a necessary condition for the emergence of a
convolution. \footnote
{Arguably the best example of  deep-inelastic scattering from a NR target
filinclusive scattering from liquid ${\rm H}_2$ molecules at low $T$.  In
the Born-Oppenheimer  there is  nearly perfect  separation of  degrees of
freedom for translations of the centers of molecules and of rotations for
frozen electronic and vibrational degrees of freedom.  The total response
is consequently a convolution \cite{you,lan}.}

\section{The total nuclear response in the PWIA.}

As stated above,
virtually all treatments of the EMC effect are based on the PWIA
applied to  the nuclear part in (\ref{a6})
\cite{btom,ross,nach,ak,dun,atw,rob,jaf2,liu,ji,mul}, thus
\begin{eqnarray}
W_{\alpha/{\rm A}}^{PWIA}(q,\nu)\equiv
\int^{\nu}_{q^2/2M_A} \,d\nu'\,W_{{\rm N/A}}^{PWIA}(q,\nu')
W_{\alpha/{\rm N}}(q,\nu-\nu'+q^2/2M)
\label{a11}
\end{eqnarray}
$W_{\NA}^{PWIA}$ above is the nuclear structure function if one neglects
in the nuclear  part  of the Hamiltonian (\ref{a1})
the interaction between a selected nucleon $'1'$ and the remaining
($A-1$)-particle core, i.e.  if
$\sum_{j<i}U(R_{ij})\rightarrow \sum_{2\le j<i}U(R_{ij})$. Excited
states in Eq. (\ref{a5}) then become
\begin{eqnarray}
|N\rangle \rightarrow |N\rangle^{PWIA}=
|(A-1)_{\lambda,-\bmP};{\bmP}\rangle,
\label{a12}
\end{eqnarray}
where $|(A-1)_{\lambda,-{\bmP} }{\bmP}
\rangle$  denotes an excited state of a moving core
with excitation and kinetic energies $E,{\cal E}^C_{\bmP}$,
and a free nucleon with momentum
${\bmP}$ and energy ${\cal E}_{\bmP}={\bmP}^2/2M$.
Substitution of (\ref{a12}) into (\ref{a5}) and the replacements
$\displaystyle \sum_N{\hspace {-0.5cm}{\int }}\rightarrow
\displaystyle \sum_{\lambda} {\hspace {-0.5cm}{\int }}
\int d{\bmP}\to\int dE d{\bmP}$ lead to
\begin{mathletters}
\label{a13}
\begin{eqnarray}
W_{{\rm N/A}}^{PWIA}(q,\nu)&=&\displaystyle
\sum_{\lambda}{\hspace {-0.5cm}{\int }}\int d{\bmP}
|{\psi}_{\lambda}({\bmP} )|^2
\delta(\nu-\Delta_{\lambda}-{\cal E}^C_{\bmP}-
{\cal E}_{\bmP +\bmq })
\label{a13a}\\
&=&\int dE\int d\bmP
S_{{\rm N/A}}(\bmP,E)
\delta (\nu -E-{\cal E}^C_{\bmP}
-{\cal E}_{\bmP+\bmq})
\label{a13b}
\end{eqnarray}
\end{mathletters}
where ${\psi}_{\lambda}({\bmP})=\langle(A-1)_
{\lambda,-{\bmP} };{\bmP}|0\rangle$,
is the overlap of the ground state of the target at rest with
an excited core state $\lambda$ and a free nucleon. The function
\begin{eqnarray}
S_{{\rm N/A}}({\bmP},E)
=\displaystyle \sum_{\lambda}{\hspace {-0.5cm}{\int }}
|{\psi}_{\lambda}(\bmP)|^2 \delta(E-\Delta_{\lambda})
\label{a14}
\end{eqnarray}
in (\ref{a13b}) is the single-particle spectral function in terms of the
separation  energy $\Delta_{\lambda}$ of a nucleon when removed from
the target ground state, and leaving the core with excitation energy $E$.
After substitution of (\ref{a13b}) into  (\ref{a11}),
the integrations of the latter over $E$, respectively $\nu'$  result in
\begin{mathletters}
\label{a15}
\begin{eqnarray}
W_{\alA}^{PWIA}(q,\nu)
&=&\int d\nu'\bigg \lbrack \int d{\bmP} S_{{\rm N/A}}
({\bmP},\nu' -{\cal E}^C_{\bmP}
-{\cal E}_{\bmP+\bmq})\bigg \rbrack
W_{\alpha/{\rm N}}(q,\nu-\nu'+{\cal E}_{\bmq})
\label{a15a}\\
&=&\int_{E_m}^{E_M} dE \int d{\bmP} S_{N/A}({\bmP},E)
W_{\alpha/{\rm N}}(q,\nu-E-{\cal E}^C_{\bmP}-{\cal E}_{\bmP+\bmq}
+{\cal E}_{\bmq}),
\label{a15b}
\end{eqnarray}
\end{mathletters}
where the  upper and lower limits $E_m, E_M$
on the $E$-integral are determined  by the  integration over $\nu'$ in
(\ref{a11})\cite{cl}.

We now discuss the above results in some detail and start with
(\ref{a15b}). Comparison with (\ref{a4}) shows that the
energy argument of $W_{\alN}$ in (\ref{a15b}) corresponds to a nucleon
with  momentum $\bmP$  and apparent off-shell energy
\begin{eqnarray}
{\cal E}^{off}({\bmP},E)=-E-{\cal E}^C_{\bmP}\ne {\cal E}_{\bmP}
\label{a16}
\end{eqnarray}
We now recall that
in  the PWIA, core and  knocked-out nucleon  are by  definition
on  their respective  energy shells (see Fig. 2). Consequently
the energy conservation expressed by the argument of the
$\delta$-function in (\ref{a13b}) prescribes a well-defined off-shell
energy  to a struck, constituent nucleon in a nucleus.

Turning to (\ref{a15a}), one  recognizes  in the  spectral function
the PWIA structure function for a semi-inclusive A$(e,e'p)$X reaction,
with knocked-out proton   momentum $\bmP+\bmq$ (Fig. 2). The
argument of the $\delta$-function in (\ref{a13b})
reflects over-all energy conservation in the PWIA and implies that
the core excitation energy
$E$ equals the missing energy $\nu-{\cal E}^C_{\bmP}-
{\cal E}_{\bmP +\bmq}$ \cite{moug}.
The integration  over the momentum of the knocked-out proton  $\bmP$
in  (\ref{a15a})  extends only over the nuclear component and the form
in brackets there (the integrated semi-inclusive response)
is the inclusive nuclear response in  the  PWIA.

Out of the two expressions (\ref{a15}) only
Eq. (\ref{a15a}) has manifestly the form (\ref{a6}), which  as recalled,
shows no trace of the motion of the a nucleon in the nucleus.
In contradistinction,  in the nuclear structure function
$W_{\NA}^{PWIA}$ in (\ref{a15a}) one deals with an off-shell
nucleon in motion.

The above is  clearly the
physical  content of  a change  of integration  variable when  going from
(\ref{a15a}) to (\ref{a15b}).
Whatever form is chosen, off-shell kinematics
is no real problem because structure functions of an
off- and an on-shell nucleon  are in a NR theory related by
(\ref{a4}) through purely kinematic quantities.\footnote
{The discussed  expression for
energy conservation  is typical for the approximation used, in  case the
PWIA.   A  different   approximation  leads  to  a   different  form  and
subsequently to  a different off-shell  energy: The very notion  does not
even emerge in the exact expression (\ref{a6}).}

As in the previous Section one may replace the energy transfer
$\nu$ by the scaling variable $y$, Eq. (\ref{a7}).
In the analog of (\ref{a11})
\begin{eqnarray}
\phi_{\alA}^{PWIA}(q,y)=\int dY\,\phi_{\NA}^{PWIA}(q,Y)
\phi_{\alN}\bigg (q,y-\frac{m}{M}Y \bigg )
\label{a17}
\end{eqnarray}
one deals with
\begin{eqnarray}
\phi^{PWIA}_{{\rm N}/{\rm A}}(q,Y)= \int dE\int d{\bmP}
S_{{\rm N/A}}({\bmP},E)
\delta \left ( Y-P_z-\frac{M}{q}(E+{\cal E}^C_{\bmP}+
{\cal E}_{\bmP})\right ),
\label{a18}
\end{eqnarray}
and again one may integrate over either
$Y$ or $E$.  As was the case with  Eqs. (\ref{a15a}),
(\ref{a15b}) above, the following two identical expressions  result
(cf. Eqs. (\ref{a13}))
\begin{mathletters}
\label{a19}
\begin{eqnarray}
\phi_{\alA}^{PWIA}(q,y)
&=&\frac {q}{M} \int dY\left [\int d{\bmP}
S_{{\rm N/A}}\left ({\bmP},\frac{q}{M}(Y-P_z)-({\cal E}^C_{\bmP}+
{\cal E}_{\bmP})\right )\right ]
\phi_{\alpha/{\rm N}}(q,y-\frac{m}{M}Y)
\label{a19a}\\
&=&\int_{E_m}^{E_M} dE\int d{\bmP}
S_{\rm {N/A}}({\bmP},E)
\phi_{\alpha/{\rm N}}\bigg (q,y-\frac{m}{M}P_z-
\frac{m}{q}(E+{\cal E}_{\bmP}^C+{\cal E}_{\bmP}) \bigg )
\label{a19b}
\end{eqnarray}
\end{mathletters}
Finally for use below,
we relate again structure functions for off- and on-shell
kinematics, using (\ref{a5}) in the $y$-variable
\begin{eqnarray}
\phi_{\alN}(q,y;\bmP,{\cal E})
=\phi_{\alN}\bigg (q,y-\frac{m}{M}P_z+\frac{m}{q}
({\cal E}-{\cal E}_{\bmP})\bigg )
\label{a20}
\end{eqnarray}
We refer to \cite{gur1} for a discussion of the asymptotic limit
and related special cases.

\section{Relativistic extension}

In order to prepare an extrapolation of the above results to the
relativistic case, it is convenient to introduce light-cone variables.
Thus   for a quark, a nucleon,  and a nuclear target
with 4-momentum $p, P$ and $P_{\rm A}$ we define  light-cone momenta
$p_{\pm}=p_0\pm p_z;\, P_{\pm}=P_0\pm P_z; \,P^A_{\pm}=M_A$, and in
addition light-cone fractions\   $p_{-}/P^{\rm A}_{-};\,
P_{-}/P_{-}^{\rm A}$. The latter we relate as usual
to a nucleon, and not to the  actual nuclear target at rest:
$x\equiv p_{-}/M;\, z\equiv P_{-}/M$.

Consider now the above light-cone  variables in the NR limit. In line
with the text after  Eqs. (7), $p_z\to y;\ \, P_z\to Y$ leading to the
following correspondences \footnote
{The sign convention in Eqs. (\ref{a21}) is in agreement
with  $y<0$, respectively $x>1$ for elastic scattering.}
\begin{mathletters}
\label{a21}
\begin{eqnarray}
x&=&\frac{p_0-p_z}{M} \,\,\rightarrow \xi=\frac{m-y}{M}
\label{a21a}\\
z&=&\frac{P_0-P_z}{M}\,\rightarrow  \zeta=\frac{M-Y}{M}
\label{a21b}
\end{eqnarray}
\end{mathletters}
The above NR momentum fractions
serve to define structure functions of on-shell targets at rest
\begin{mathletters}
\label{a22}
\begin{eqnarray}
f_{\alpha /{\rm N}}(q,\eta)
&=& M\phi_{\alpha/{\rm N}}(q,m-M\eta)
\label{a22a}\\
f_{\rm {N/A}}(q,\eta) &=& M\phi_{{\rm N}/{\rm A}}(q,M-M\eta)
\label{a22b}\\
f_{\alpha /{\rm A}}(q,\eta)&=&
M\phi_{\alpha/{\rm A}}(q,m-M\eta)
\label{a22c}
\end{eqnarray}
\end{mathletters}
Analogous to (\ref{a4}) and (\ref{a20}) we can express the structure
function for a moving off-shell nucleon by means of the same at rest
\begin{eqnarray}
f_{\alN}(q,\xi;\bmP,{\cal E})
= M\phi_{\alpha/{\rm N}}\left ( q,y -\frac{m}{M}P_z
-\frac{m}{q}\Delta{\cal E}\right )\cong f_{\alN}
\left (q,\frac{M\xi}{M(1-\Delta{\cal E}/q)-P_z}\right )
\label{a23}
\end{eqnarray}
Above $\Delta{\cal E}={\cal E}_{\bmP}-{\cal E}$
the off-shell energy shift, and in the last step in (\ref{a23})
we used the approximation
$$y-\frac{m}{M}P_z-\frac{m}{q}\Delta{\cal E}\cong m-
\frac {M(m-y)}{M(1-\Delta{\cal E}/q))-P_z}$$

Just as  Eqs. (\ref{a6}), (\ref{a10})  express the nuclear response  as a
convolution in  $\nu$ and  a  NR scaling  variable $y$, one can give a
third representation in terms  of the
variables  $\eta$. Substitution of (\ref{a21}) into  (\ref{a10}), the use
of the definitions (\ref{a22})
and the approximation $m-y+(m/M)Y\cong M(\xi/\zeta)$
lead to
\begin{mathletters}
\label{a24}
\begin{eqnarray}
f_{\alpha/{\rm A}}(q,\xi)&=&
\int d\zeta
f_{{\rm N}/{\rm A}}\left (q,\zeta \right )
f_{\alpha/{\rm N}} \bigg(q,\xi- \frac{m}{M}\zeta+\frac{m}{M}\bigg)
\label{a24a}\\
&\cong& \int d\zeta f_{{\rm N}/{\rm A}}
\left(q,\zeta\right) f_{\alpha/{\rm N}} \bigg (q,\frac{\xi}{\zeta}
\bigg)
\label{a24b}
\end{eqnarray}
\end{mathletters}

All results till (\ref{a24a}) are {\em exact}. In particular
the latter  expression contains in the $q$-dependent nuclear component
$f_{\NA}$ the full  nucleon-nucleon dynamics, by which token it  is
equivalent to the nucleon analog $f_{\alN}$.

Eqs. (\ref{a24}) are a direct consequence of a strictly
NR cluster model, which has of course no link to
relativistic  dynamics for the
constituents. Notwithstanding, the above result has clear features of
relativistic expressions for the total nuclear structure function
in the literature.  This observation invites to dissociate the
outcome (\ref{a24}) from the underlying model and to attempt an
extension into the relativistic domain.
\footnote  {Such a procedure is  by no means exceptional.  A striking
example is  Glauber theory  for the scattering  from a  composite target,
which  has been  derived from  a  NR theory  with real  pair-interactions
between  constituents.  The  final result,  free of  any reference  to NR
elements in general,  and classical interactions in  particular, has been
postulated  to hold,  and  has with  great success  been  applied in  the
relativistic  regime  \cite{glaub}.} This we do, first  reverting  to the
relativistic light-cone fractions in Eq. (\ref{a21}).  Then with
$Q^2=q^2-\nu^2$ and $x^{Bj}=Q^2/2M\nu$, the replacements
$$q=Q\sqrt{1+(Q/2Mx^{Bj})^2};\,\,\,  \xi\to x^{Bj}$$
lead to
\begin{eqnarray}
f_{\alA}(q,\xi) \to F_{\alA}(x^{Bj},Q^2)
\label{a25}
\end{eqnarray}
and to similar correspondences for other structure functions.
We now conjecture that in the relativistic regime
\begin{eqnarray}
F_{\alA}(x,Q^2)=\int^{M_A/M}_x \,dzF_{\NA}(z,Q^2)
F_{\alN} \bigg(\frac{x}{z}, Q^2\bigg),
\label{a26}
\end{eqnarray}
with $x\to x^{Bj}$ for large, finite $Q^2$
is the relativistic analog  of  Eq. (\ref{a24b}).

We now follow the same line as in the NR treatment and consider the
PWIA for the nuclear response. We assume that nuclear
states can be described
non-relativistically and that the same holds for derived quantities
$\psi_{\lambda}$ or the spectral function $S_{{\rm N/A}}$,
Eq. (\ref{a15}). The sole relativistic aspect retained
in this soft component is the kinematics of the knocked-out nucleon.
We thus write for the structure function $W_{\NA}(q,\nu)$ in the PWIA
with relativistic kinematics (Fig. 2)
\begin{eqnarray}
W_{\NA}^{PWIA}(q,\nu)=
\int dE\int d^4P\,2M|\psi({\bmP},E)|^2
\delta\left [ (P_0+\nu )^2-({\bmP+\bmq})^2-M^2\right ]
\delta (P_0-P_0^{off})
\label{a27}
\end{eqnarray}
As in the NR case (cf. (\ref{a13a})), the energy argument above
is determined by over-all energy conservation which
determines the off-shell total energy of the struck nucleon
\begin{eqnarray}
P_0^{off}({\bmP},E)=M_A-[(M_{A-1}+E)^2+{\bmP}^2]^{1/2}
\cong M-E-{\cal E}^C_{\bmP}
\label{a28}
\end{eqnarray}
Then using Eqs. (\ref{a26}) and  (\ref{a27}),
the total nuclear structure function in the PWIA
(see Fig. 1) becomes
\begin{mathletters}
\label{a29}
\begin{eqnarray}
F^{PWIA}_{\alpha/{\rm A}}(x,Q^2) &=&
\int^{M_A/M}_x\,dz F^{PWIA}_{\NA}(z,Q^2)
F_{\alN} \bigg(\frac{x}{z},Q^2 \bigg)
\label{a29a}\\
F_{\NA}^{PWIA}(z,Q^2)&\equiv&
\nu W_{\NA}^{PWIA}(q,\nu)
\label{a29b}
\end{eqnarray}
\end{mathletters}
Consider now large $q^2\gg M^2+{\bmP}^2$ for which
\begin{eqnarray}
\lbrack(\bmP+\bmq)^2+M^2\rbrack ^{1/2}
\rightarrow q+P_z+{\cal O}({\bmP}^2/q )
\label{a30}
\end{eqnarray}
When used in (\ref{a27}),
Eq. (\ref{a29b}) becomes (cf. Eqs. (\ref{a13}), (\ref{a18}))
\begin{eqnarray}
F_{\NA}^{PWIA}(z) \to \int dE \int d{\bmP}
S_{\NA}({\bmP},E)\delta (z-\frac{P^{off}_0-P_z}{M}\bigg )
\label{a31}
\end{eqnarray}
Upon  substitution of (\ref{a31}) into (\ref{a29a}) one may again
integrate  over $E$ or $z$ with the results
\begin{mathletters}
\label{a32}
\begin{eqnarray}
F_{\alA}^{PWIA}(x,Q^2) &=& M\int^{M_A/M}_x\,dz \bigg \lbrack
\int d{\bmP}S_{\NA}\left ({\bmP}, M(1-z)-P_z-{\cal E}^C_{\bmP}\right )
\bigg \rbrack F_{\alN}\bigg(\frac{x}{z},Q^2\bigg )
\label{a32a}\\
&=&\int^{E_M}_{E_m} dE \int d{\bmP}
S_{\NA}({\bmP},E)F_{\alN}
\left ( \frac{x}{(P_0^{off}({\bmP},E)-P_z)/M},Q^2\right )
\label{a32b}
\end{eqnarray}
\end{mathletters}

All  remarks made after Eqs.  (\ref{a16}) apply here as well.
Eq. (\ref{a32a}) accounts implicitly for the motion of a target nucleon
in which $F_{\alN}$ has no part and is  a structure
function for a nucleon at rest with on-shell kinematics.
In contrast, the alternative expression (\ref{a32b})
contains a loop integral
over $(E,{\bmP})$ embracing both components $S_{\NA}$ and $F_{\alN}$.
It is close to the standard PWIA expression, where usually explicit
mention is made of a nucleon structure function
$F_{\alN}(x,Q^2;\bmP,P_0)$ (see for instance \cite{mel,kul}.
Its identity with (\ref{a32b})
is established by means of the following relativistic analog of
(\ref{a23})
\begin{eqnarray}
F_{\alN}(x,Q^2;\bmP,P_0^{off})\rightarrow  F_{\alN}
\bigg(\frac{x}{(P_0^{off}({\bmP},E)-P_z)/M},Q^2 \bigg )
\label{a33}
\end{eqnarray}
Incidentally, in  view of (\ref{a30})  and the relative smallness  of the
recoil kinetic energy of the core  in (\ref{a28}), the dependence of the
integrand in (\ref{a32b}) on $\bmP_{\perp}$ resides in practice only in
$S_{\NA}$.

\section{Discussion}

We  have  discussed above  a  non-relativistic  quark-cluster model  with
triplets of spin-0 quarks, confined  to bags, which amongst them interact
by means of effective nucleon-nucleon forces.   In such a model one shows
that, for arbitrary $q$, the  total nuclear response $W_{\alA}(q,\nu)$ is
a convolution  of responses for a  nucleus made up of  point-nucleons and
for  a  nucleon  bag.   With  no manifest  reference  to  the  underlying
potential model we postulated an  extension into the relativistic regime.
The result Eq. (\ref{a26}) formally contains the full nuclear interaction
in $W_{\NA}(q,\nu)$,  i.e. incorporates Final State  Interactions between
the core  and the  nucleon and  which by definition  is neglected  in the
PWIA.  Moreover, separated nucleonic and sub-nucleonic degrees of freedom
clearly leave no  room for rescaling or swelling of  nucleons as a result
of embedding in a nucleus.

The   above-mentioned   effective   nucleon-nucleon  forces   cannot   by
construction  describe  a detailed  interaction  of  quarks in  different
nucleons, but only  their global effect.  For  example interactions there
between the di-quark and of the  recoiling quark debris with a given core
nucleon  are  parts of  an  effective  $NN$  interaction.  In  the  above
restricted  sense  $F_{\NA}$ in  the  model  result (\ref{a26})  formally
contains  all Final  State  Interaction contributions  ($'$ higher  order
twist$'$).

A  sufficient  condition  for  (\ref{a26}) to  hold  is  separability  of
sub-hadronic from nuclear  degrees of freedom.  It is  also calculable in
principle.  There  actually is  no need  to specify  the dynamics  of the
constituents, or  even to  be explicit  about their  nature and  the hard
portion  of the  total structure  function $F_{\alN}$  can be  taken from
data.  In contradistinction the soft part of the total structure function
$F_{\NA}$ is presently not accessible  \cite{gur1} and one needs explicit
nuclear dynamics for an actual calculation.

A model with extreme simplifications as the one discussed above obviously
misses  essentials  of a relativistic  description
of inclusive scattering from  a nucleus. For  one, spin
effects complicate a simple convolution of the form (\ref{a26}).  We now
discuss two basic elements which enable an evaluation of
$F_{\alA}$, namely quark  clustering in nucleon bags and  their mutual
interaction.  The  latter leads to FSI  between the core
and  nucleon   debris,  independent  of   the  state  of   excitation  or
hadronization.  Essentially  the same assumptions  have been made  in far
more  sophisticated approaches,  where  triplets of  quarks  in MIT  bags
couple  to   mesons  \cite{sait1}.   Those  couplings   induce  effective
interactions between  bags or nucleons,  mediated by the exchange  of the
same mesons. Clearly the
notion of bags in a larger  assembly of quarks
is tantamount to clustering, while  $qqM$ couplings induce effective $NN$
interactions mediated by the same mesons.  Also in an application of that
model to the total nuclear structure  function, again only the bag ground
state appears involved  \cite{sait2}. As in the quark  cluster model, the
$qqM$ coupling model completely disregards the state of excitation of the
bag.

In the work  by Melnitchouk $et\,al$  one can find another
attempt to  go beyond
the PWIA,  in case by estimating  the FSI between the  di-quark debris of
the struck nucleon  and the core.  We already remarked  that in a cluster
model the effective $NN$ interaction  combines the
interaction of a core nucleon with  the debris of the struck
nucleon , i.e.  a  di-quark $and$ the recoiling quark.

In summary we discussed a quark cluster model which provides answers
on some cogent questions, relevant for  a treatment of the EMC
effect. It is certainly of interest to see those questions addressed
and handled in more sophisticated theories.

{\bf Figure Captions.}

Fig. 1. Amplitude appearing in the total  nuclear   structure  function
$W_{\alA}$  in  PWIA.

Fig. 2.  Amplitude, displaying
kinematics for nuclear structure function $W_{\NA}$ in PWIA.
Crosses mark on-shell particles.

\end{document}